\documentclass[journal]{IEEEtran}
\usepackage{amsmath,epsfig}
\usepackage{graphicx}
\usepackage{siunitx}
\usepackage{subcaption}
\usepackage{xcolor}
\usepackage{stfloats}

\usepackage{bm} 
\usepackage{cite}
\usepackage{soul}
\usepackage{booktabs}
\usepackage{multirow}
\usepackage[hidelinks]{hyperref}
\usepackage{url} 

\makeatletter
\g@addto@macro{\UrlBreaks}{\do\-}
\makeatother

\ifCLASSINFOpdf
\else
\fi
\begin{document}
%

\title{Quantifying System-Level Performance Bottlenecks in Hyperspectral Target Detection}


%
%
%

\author{Chase~Ca\~{n}as~and~John~P.~Kerekes
\thanks{C. Ca\~{n}as and J.P. Kerekes are with the Chester F. Carlson Center
for Imaging Science, Rochester Institute of Technology, Rochester,
NY, 14623 USA e-mail: chasecanas@gmail.com}
\thanks{This work was supported in part by NGA University Research Initiative grant HM04762110005. Approved for public release, NGA-U-2025-02112}}

%
%


\markboth{}
{Ca\~{n}as \MakeLowercase{\textit{et al.}}: System Limits in Target Detection}

%



\maketitle



\begin{abstract}

Discerning core performance bottlenecks in hyperspectral target detection is highly challenging given the complex interplay between sub-systems of the imaging chain (scene, atmosphere, sensor and processing). Prior sensitivity and limitation studies offer narrow insights applicable to specific system configurations. To address the need for broader generalization, an analysis framework was constructed to quantify limitations from large ensembles of model scenarios. Key components of the framework include the use of canonical scene representations and the log-weighted area under the ROC curve (wAUC), a metric well-suited for rare target detection. By analyzing thousands of scenarios, parametric thresholds were quantified where detection performance degrades, corresponding to ``knees" in performance curves. A knee significance score (KSS) is introduced to assess the relative importance of these thresholds. The framework an airborne VNIR sensor implanted in 28,000 scenarios using seven system parameters spanning the scene, atmosphere and sensor. Results suggest that aerosol visibility in the 2--6 km range is the dominant limiter in detection, followed by subpixel target percentages of 15--35\%, and background clutter modeled using a t-distribution with 3--5 degrees of freedom.

\end{abstract}


%
\IEEEpeerreviewmaketitle

%
%
%
%


 



\begin{figure*}[b!]
    \centering
    \includegraphics[width=0.85\linewidth]{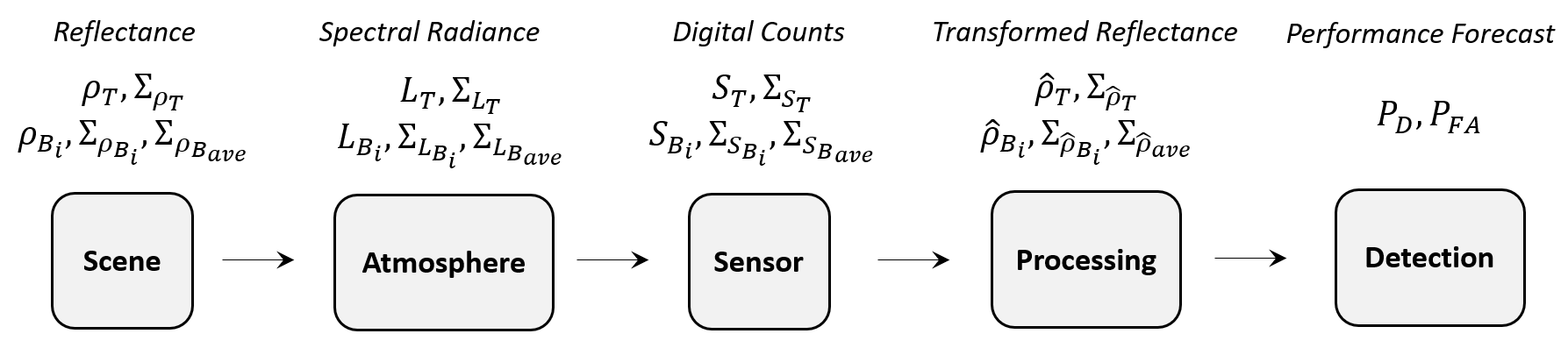}
    \caption{Hyperspectral systems model (FASSP) transforms first- and second-order statistics through the imaging chain.}
    \label{fig:fassp_flowchart}
\end{figure*}

\section{Introduction}

\IEEEPARstart{S}{ystems} limitations in optical hyperspectral target detection arise from complex interactions across the imaging chain (scene, atmosphere, sensor, processing). Despite decades of significant progress in algorithm development for hyperspectral target detection \cite{manolakis_detection_2002,kwon_comparative_2006,jin_comparative_2009,nasrabadi_hyperspectral_2014}, open questions remain regarding the fundamental detection limitations from an end-to-end systems perspective. Understanding how detection performance degrades with variations in scene composition, atmospheric conditions, sensor specifications, and data processing approaches is crucial for forecasting limitations in operational scenarios and informing optimal sensor design.

While a handful of researchers have developed systems-based models for hyperspectral target detection \cite{kerekes_hyperspectral_2003, bernhardt_hyperspectral_2007}, the application of these models to analyze end-to-end limitations across an extensive range of scenarios remains underdeveloped. Prior research has explored system-level limitations arising from parameters such as subpixel fraction, sensor noise, and scene complexity, but such studies have been constrained to specific scenarios \cite{kerekes_model-based_2006,kerekes_exploring_2011}. Furthermore, the broader literature on hyperspectral limitations has primarily concentrated on isolated components of the imaging chain. This includes studies on impacts of atmospheric and illumination effects \cite{ientilucci_impact_2009,ientilucci_share_2013,sundberg_impact_2018}, sensor and instrumental effects \cite{casey_misregistration_2009,morman_hyperspectral_2015,han_optimizing_2020}, background variability \cite{kasen_target_2004,manolakis_realistic_2005}, target percentage \cite{kerekes_share_2013,gross_assessment_2015}, and processing algorithms \cite{dipietro_performance_2010,gross_determination_2016}. While these compartmentalized approaches are valuable, they offer limited insight into how performance limitations emerge across the full imaging chain of remote sensing systems.

Despite the growing operational relevance of hyperspectral sensors, a holistic understanding of generalized detection limitations accounting for interactions among the full imaging chain remains incomplete. To address this gap, we introduce a scene generation and analysis framework for characterizing performance bottlenecks in hyperspectral target detection. Novel research contributions include the use of a log-weighted AUC (wAUC) combined with a proposed knee significance score (KSS) for ranking performance-limiting parameters and associated thresholds. To demonstrate the framework, a case study for a VNIR airborne sensor is presented using an ensemble of 28,000 scenarios constructed from combinatorial values of system parameters. Data generation of canonical scenes (forest, urban, rural, desert) is orchestrated by adopting a prior statistical systems model tailored for subpixel target detection (FASSP), as described in the following subsection.

\subsection{Statistical Systems Model for Subpixel Target Detection}
\label{sec:fassp}

The Forecasting and Analysis of Spectroradiometric System Performance (FASSP) is a statistical systems model for examining limitations in hyperspectral target detection \cite{kerekes_spectral_2002}. This previously validated model defines a detection scenario by incorporating parameters across all stages of the remote sensing system (scene, atmosphere, sensor, processing). The model operates by transforming first- and second-order spectral statistics (mean vectors and covariance matrices derived from empirical measurements) for each target and background class through the imaging chain. As outlined in Fig.~\ref{fig:fassp_flowchart}, these statistics are sequentially transformed across domains: surface reflectance, at-sensor radiance, digital counts, and calibrated reflectance.

Atmospheric parameters are defined via MODTRAN \cite{berk_modtran5_2005}, with a physics-based sensor model simulating noise and digitization. Optional post-processing includes atmospheric compensation (e.g., empirical line method) and feature selection (e.g., band selection, principal component analysis). Finally, the transformed statistics are passed to a detector algorithm (e.g., matched filter), which outputs a Receiver Operating Characteristic (ROC) curve as the performance metric.

\subsection{Overview}

This research constructs an end-to-end framework for quantifying limitations in hyperspectral target detection, applied to an airborne case study across seven system-level remote sensing parameters. Methodologically, there are two primary components of the framework:
\begin{itemize}
    \item A \textit{model-based} workflow for statistical data initialization of scenarios.
    \item An \textit{analytical} workflow for conducting trade studies to quantify system-level limitations.
\end{itemize}

In Sec.~\ref{sec:model_framework} the model-based approach is outlined for data initialization of large-scale parametric trade studies. Sec.~\ref{sec:analytical_framework} outlines the analytical methodology for identifying specific parametric thresholds where detection sharply degrades. In Sec.~\ref{sec:case_study}, a case study applies these methodologies to analyze limitations across seven remote sensing parameters. Sec.~\ref{sec:discussion} interprets the operational implications of the case study results. Finally, Sec.~\ref{sec:conclusion} summarizes the research and offers recommendations for future research.

\section{Methodology to Initialize Model-Based Scenarios}
\label{sec:model_framework}

We introduce a traceable, model-based methodology for initializing trade studies across thousands of scenarios, using a reconfigured version of the statistical systems model FASSP. The reconfiguration, developed by the RIT Digital Imaging and Remote Sensing Image Generation Model (DIRSIG) team, has significantly reduced computational runtimes, enhancing FASSP's utility for large-scale trade studies. FASSP is integrated as a subroutine within a broader model-based analytical wrapper for investigating system-level limitations.

\begin{figure}[h!]
    \centering
    \includegraphics[width=\linewidth]{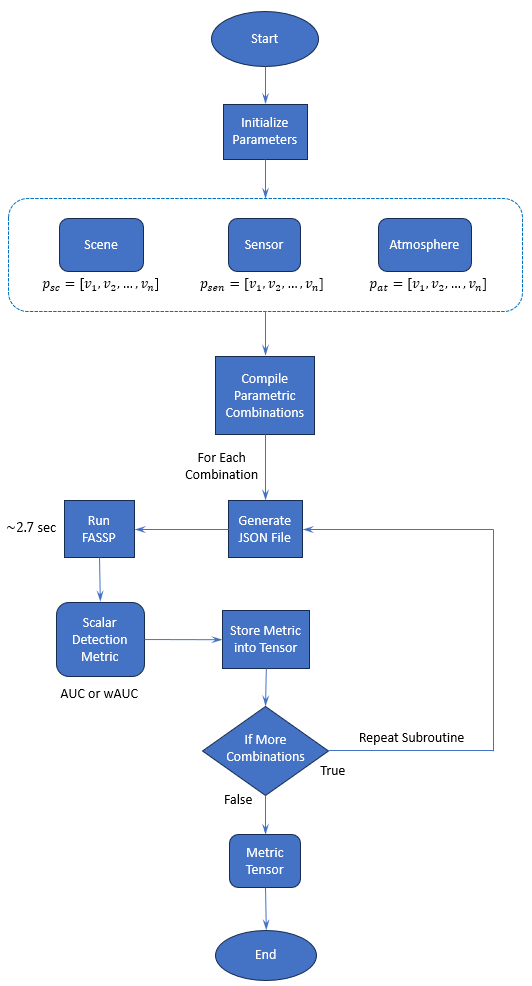}
    \caption{Computational pipeline to generate and store output detection metrics from FASSP into a tensor (high-dimensional array), given unique sets of input parameters.}
    \label{fig:workflow}
\end{figure}

The model-based workflow begins by initializing linear arrays of system parameters, where unique sets of combinatorial values define a large ensemble of detection scenarios. For each scenario, a FASSP subroutine ($\sim$2.7 seconds) using MODTRAN is executed on a high-powered Linux machine (386 GB memory and 48 CPU cores) at the Chester F. Carlson Center for Imaging Science. A detection scalar output (e.g. AUC, wAUC) for each scenario is stored in a high-dimensional tensor for subsequent analysis, as detailed in Sec.~\ref{sec:analytical_framework}. The pipeline is summarized in Fig.~\ref{fig:workflow}.

There are two critical sub-processes within this data initialization workflow:

\begin{itemize}
    \item Constructing a \textit{statistical} representation of canonical scenes with adjustable background clutter.
    \item Computing a \textit{log-weighted} area under the ROC curve metric for rare target detection studies.
\end{itemize}

The following subsections expand on these processes, detailing their role in modeling diverse scenes and evaluating target detection performance under realistic conditions.

\begin{figure*}[b!]
    \centering
    \includegraphics[width=0.9\linewidth]{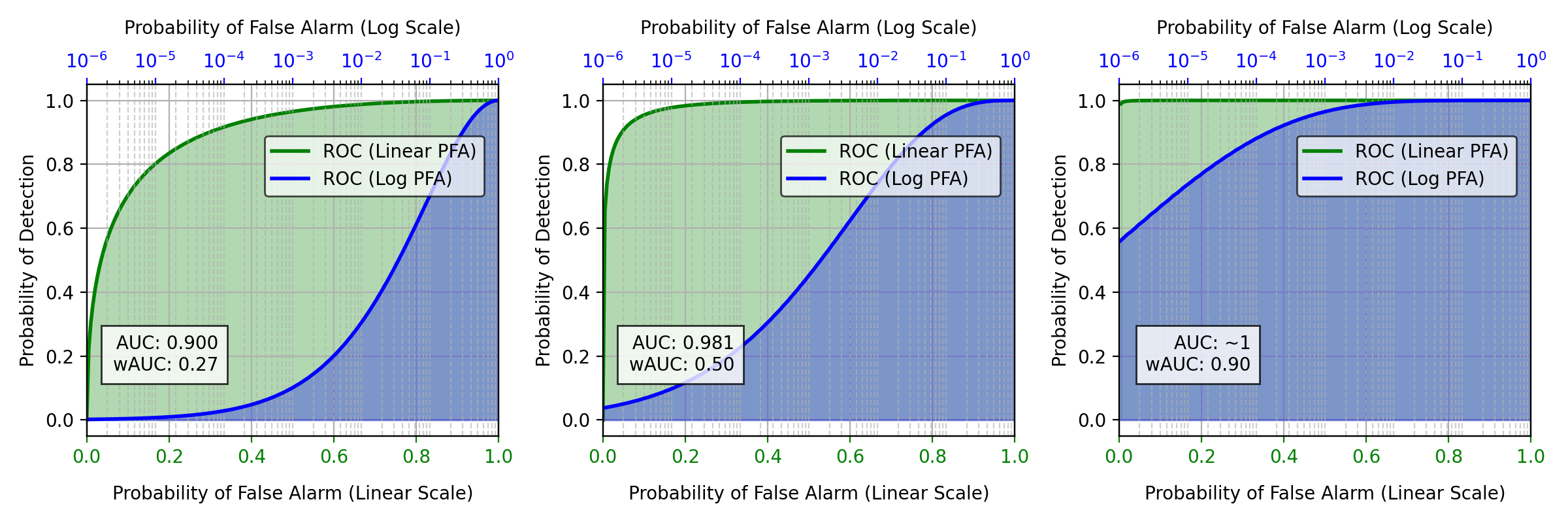}
    \caption{Comparison of wAUC and AUC for three example ROC curves representing different levels of detection performance. While the AUC metric varies only from 0.9 to 1.0, the wAUC provides a greater dynamic range varying from 0.27 to 0.90 demonstrating greater numerical sensitivity to detection performance.}
    \label{fig:wAUC_vs_AUC}
\end{figure*}

\subsection{Statistical Representation of Canonical Scenes}
\label{ssec:qualitative_scenes}

Hyperspectral scenes contain vast amounts of spatial and spectral information from various material classes (e.g., grass, soil, gravel, and trees). For any given scene, the possible combination of materials and their proportions are virtually limitless. To overcome this complexity, we generate qualitative \textit{canonical scenes} (e.g., forest, urban, rural, desert) that represent Earth environments in operational scenarios.

Statistical representations of canonical scenes are generated in accordance with the FASSP model~\cite{kerekes_hyperspectral_2003}. Unlike expensive, Monte Carlo approaches that simulate large numbers of random pixels to estimate statistics, we represent scenes analytically from spectral libraries of individual background classes. We establish a generalized scene using a set of material classes and area fractions representative of an Earth environment. Next, average background statistics are computed from the composite of spectral statistics.

The mean background spectrum across all classes is calculated as a weighted average of the associated area fractions. The average background spectral covariance matrix of the mixture is computed using the law of total covariance:

\begin{equation}
	\bm{\Sigma}_{ave} = \sum_i w_i(\bm{\Sigma}_i + \bm{\rho}_i\bm{\rho}_i^T) - \bm{\rho}_{ave}\bm{\rho}_{ave}^T
	\label{eq:avg_back_cov}
\end{equation}

where $w_i$ are normalized weights of each background class, $\bm{\Sigma}_i$ are individual covariance matrices, and $\bm{\rho}_{ave}=\sum_i w_i\bm{\rho}_i$ is the average background vector. The outer product term $\bm{\rho}_{ave}\bm{\rho}_{ave}^T$ captures variability between class means. The formulation accounts for both intra-variability within each distribution and inter-variability between the means of distributions.

Empirical observations indicate that matched filter output distributions are often heavy-tailed, especially for large scenes \cite{manolakis_realistic_2005}. To capture this additional variability, we model background clutter using a Student’s t-distribution applied to matched filter outputs, with a tunable degree-of-freedom (t-DOF) parameter. The t-DOF controls the tail heaviness of the distribution: as it increases, the distribution approaches Gaussian behavior, reducing the probability of extreme background scores. For an expanded discussion on t-DOF selection and its effect on scene modeling, see Ca\~{n}as \cite{canas_end--end_2025}.

\subsection{Log-Weighted Area Under the ROC Curve Detection Metric}

In hyperspectral target detection, the ROC curve is a widely adopted metric to benchmark detection performance. However, ROC curves are vector-valued functions that contain an abundance of information: a set of detection and corresponding false alarm rates $\{P_D,P_{FA}\}_{i=1}^n$. When evaluating hundreds or thousands of scenarios in large-scale trade studies, the utility of full ROC curves reduces and it becomes more practical to adopt a \textit{scalar} detection metric for performance assessment.

Two commonly used scalar performance metrics include the area under the ROC curve (AUC) and the detection probability at a fixed false alarm rate ($P_D|P_{FA}$). The AUC provides a comprehensive measure of performance across all operating points, but lacks sensitivity in the low false alarm rate region (e.g. $P_{FA}<0.001$), which is often critical in rare target detection scenarios. In contrast, the fixed rate detection metric ($P_D|P_{FA}$) enables focused evaluation at small false alarm rates, but it does not reflect performance across the broader range of operating conditions.

To address these shortcomings, we adopt the \textit{log-weighted area under the ROC curve} (wAUC), based on prior work including the partial AUC (pAUC) \cite{clark_managing_2008} and the logarithmically scaled AUC (logAUC) formulations developed for virtual screening \cite{mysinger_rapid_2010}. Also applied in hyperspectral anomaly detection \cite{borghys_comparative_2012}, this metric preserves the global interpretability of the standard AUC while placing greater emphasis on detection performance at low false alarm rates, regions critical for detecting rare or anomalous objects. When the false alarm rate axis of a ROC curve is represented on a logarithmic scale, the wAUC can be computed using rectangular integration as follows:

{\small
\begin{equation}
	\text{wAUC} = \frac{\sum_{i=1}^{N-1} P_{D_i} \cdot (\log P_{FA_{i+1}} - \log P_{FA_i})}{\log P_{FA_N} - \log P_{FA_1}}
\end{equation}
}

which is unit-normalized across the tested range of false alarms $\{P_{FA}\}_{i=1}^n$. An alternative formalism using trapezoidal integration is constructed by substituting $(P_{D_i} + P_{D_{i+1}}) / 2$ for $P_{D_i}$. Fig.~\ref{fig:wAUC_vs_AUC} illustrates the greater numerical sensitivity of the wAUC metric compared to AUC for three sets of identical ROC curves representing a range of detection performance.

In context to the data pipeline in Fig.~\ref{fig:workflow}, system trade studies are initialized by generating and storing a wAUC detection metric for each scenario, defined by a unique combination of parametric values. Across all scenarios, wAUC values are stored in a tensor with dimensions equal to the number of parameters in the trade study. Projections of high dimensional tensors onto three-dimensions form a \textit{detection cube}.

\begin{figure}[h!]
    \centering
    \includegraphics[width=0.7\linewidth]{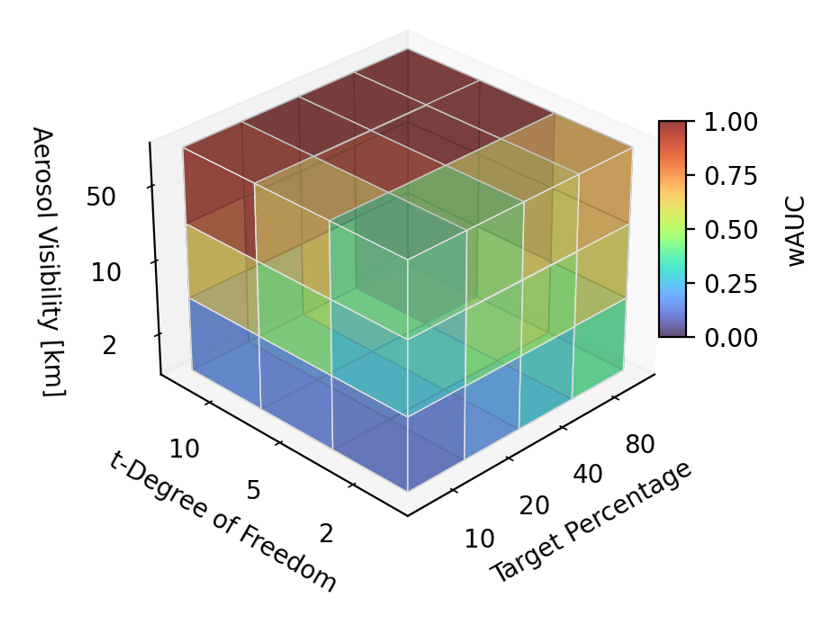}
    \caption{Detection cube (tensor) for an example trade study between three system parameters: target percentage, t-degree of freedom, and aerosol visibility.}
    \label{fig:detection_cube}
\end{figure}

Detection cubes facilitate data visualization and support inferences derived between system parameters and detection performance. An example is shown in Fig.~\ref{fig:detection_cube}. The \textit{red} region corresponds to system configurations with \textit{high} detection performance, while the \textit{blue} region indicates \textit{low} performance. In this demonstration, increased target percentage (scene), t-degree of freedom (scene), and aerosol visibility (atmosphere) are positively correlated with improved detection performance. These trends intuitively align with expectations.

The following section builds on these observations by identifying parametric thresholds, or bottlenecks, where critical transitions in detection performance occur. A novel metric is introduced to quantify their significance as indicators of system-level limitations.

\section{Methodology to Quantify System-Level Limitations}
\label{sec:analytical_framework}

After generating ensembles of model-based scenarios with detection performance outputs (wAUC) stored in high-dimensional tensors, an analytical methodology is performed to quantify system-level limitations. For each canonical scene, a two-step approach is applied to evaluate both global correlation and local limitations in detection using sequential coarse and fine sampling densities of parametric combinations. \textit{Coarse} sampling identifies and ranks parameters that are most sensitive to detection performance (wAUC). While, \textit{fine} sampling is used to identify specific parametric thresholds (among the top-ranked parameters) when detection is limited. Together, this approach offers a multi-scale understanding of system-level bottlenecks. An overview of the approach is as follows:

\begin{itemize}
    \item A \textit{global} sensitivity analysis ranks system parameters based on their overall influence on detection performance.
    \item A \textit{local} limitations analysis identifies thresholds, among the top-ranked parameters, where detection performance sharply degrades.
\end{itemize}

The following subsections expand on these analytical steps, detailing methods for ranking parameter sensitivities and identifying performance limitations across the system trade space.

\subsection{Sensitivity Analysis}

The \textit{global} sensitivity analysis uses a correlation-based approach to quantify how system parameters influence detection performance (wAUC) across a broad range of scenarios. First, a wAUC metric tensor is generated (see Sec.~\ref{sec:model_framework}) by coarsely sampling parameters of interest for each qualitative scene. Next, Spearman’s rank correlation coefficients are computed between the wAUCs and corresponding parameter values to capture the monotonic direction and strength of the inherently \textit{nonlinear} relationships common in remote sensing systems. Parameters are then ranked by correlation strength to identify those with the greatest influence. The top-ranked parameters, most sensitive to performance changes, are further analyzed in the following section to derive system-level limitations.

\subsection{Limitations Analysis}

The \textit{local} limitations analysis employs a curvature-based approach to characterize thresholds, among the top-ranked parameters, where detection performance sharply degrades. First, a new wAUC metric tensor is generated by finely resampling the most influential parameters within a narrowed range of values for each qualitative scene. This sampling can be performed iteratively, centering the parameter space around anticipated degradation regions to ensure sufficient resolution near performance-limiting thresholds. Next, univariate detection curves are extracted from the refined wAUC tensor along each parameter axis. Parameter thresholds are identified using a knee-detection algorithm, with their relative importance quantified by a novel proposed \textit{knee significance score} (KSS). The following subsections expand on these methodologies.

\subsubsection{Extraction of Univariate Detection Curves}

Univariate detection curves are extracted from the wAUC metric tensor to isolate the effects of individual system parameters. For each canonical scene, the wAUC tensor has dimensionality equal to the number of system parameters in the trade study. Detection curves are extracted ``row-by-row" along each parameter axis, holding all other parameters fixed.

For example, a 3-dimensional tensor with size $10\times10\times10$, yields three sets of 100 univariate detection curves. Each set captures how detection performance varies across the 10 sampled values of a single parameter. These curves provide the foundation for identifying performance-limiting thresholds in the subsequent analysis.

\subsubsection{Identification of Parametric Thresholds using the Kneedle Algorithm}

\begin{figure}[t!]
	\centering
	\begin{subfigure}{0.7\linewidth}
		\includegraphics[width=\linewidth]{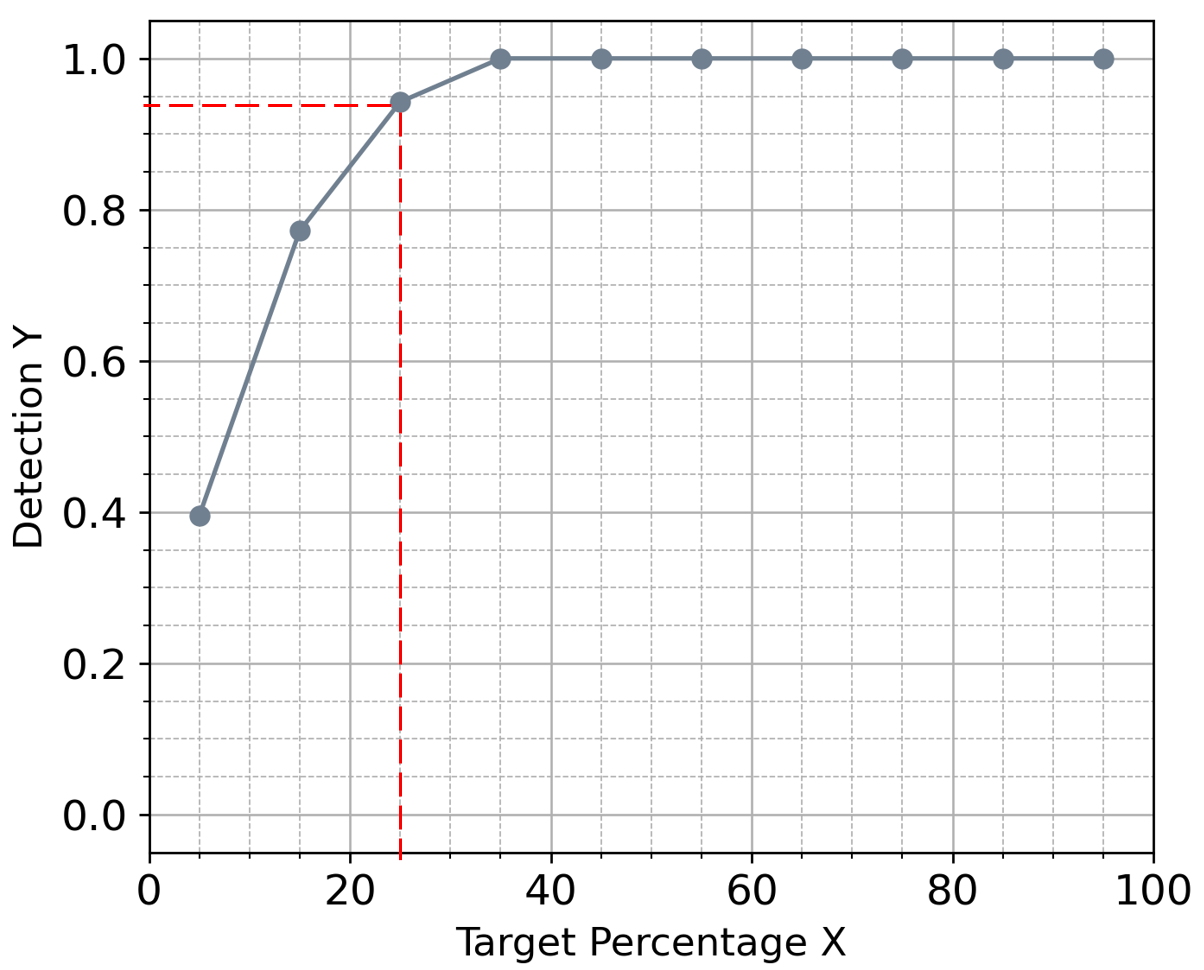}
		\caption{Raw space}	
	\end{subfigure}

        \vspace{1em}
    
	\begin{subfigure}{0.7\linewidth}
		\includegraphics[width=\linewidth]{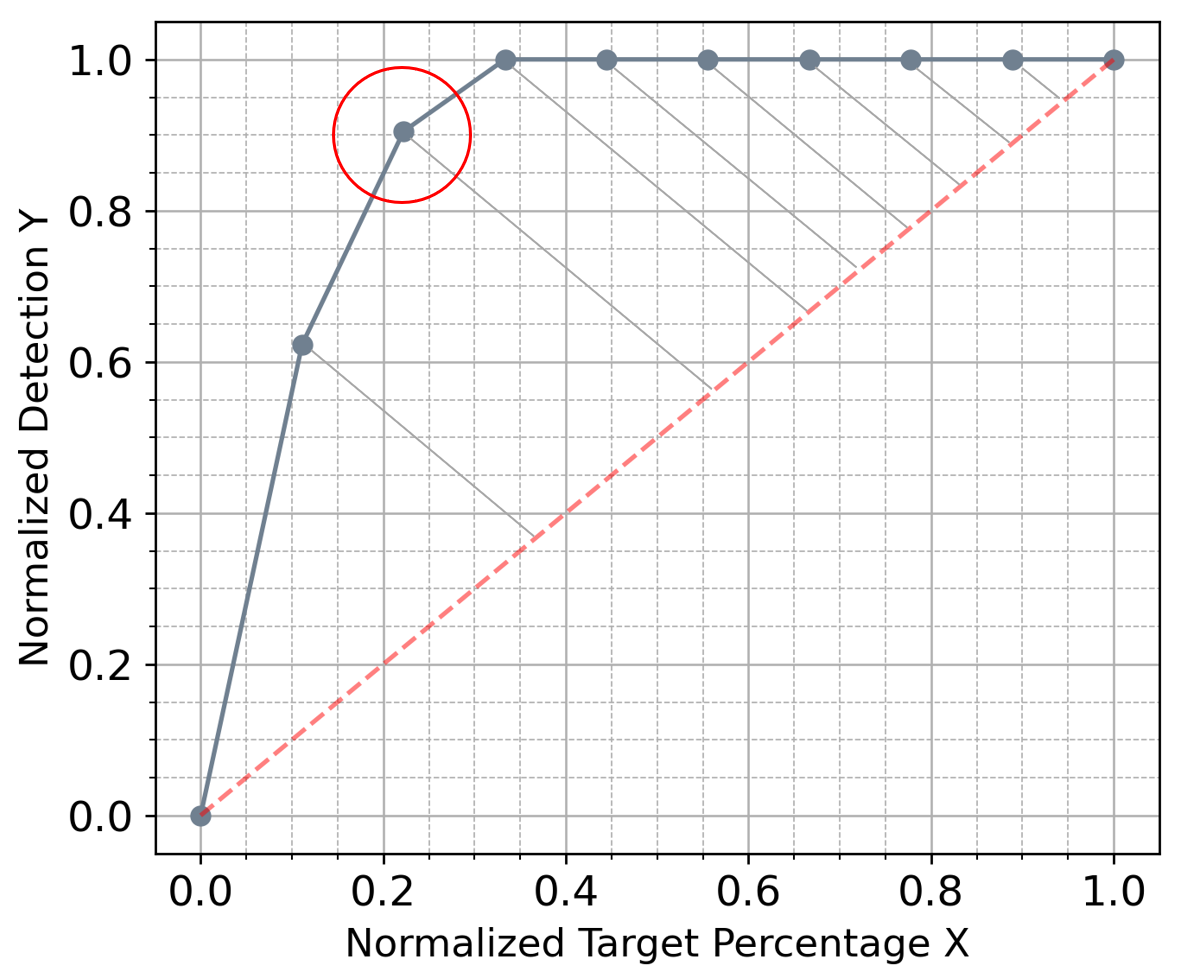}
		\caption{Normalized space}	
	\end{subfigure}

        \vspace{0.5em}

        \caption{(a) Example univariate detection curve in raw unit space, with a target percentage threshold of 25\% identified using (b) the Kneedle algorithm, which locates the point of maximum perpendicular distance from the endpoint line in normalized space.}
	\label{fig:kneedle}
\end{figure}

To identify performance-limiting thresholds within the univariate detection curves, a curvature-based method is employed to locate the points of rapid decline. The Kneedle algorithm detects such ``knees" or ``elbows" by identifying the point of maximum perpendicular distance from a straight line connecting the curve's endpoints \cite{satopaa_finding_2011}. Rather than relying on derivatives or statistical thresholds, Kneedle operates on the geometric shape of the curve, making it well-suited for noisy or non-linear functions. The algorithm is also robust to scale differences across parameters, as it operates in a normalized vector space. In this context, a knee point represents a parametric threshold and corresponding wAUC value where detection performance begins to sharply degrade. An example using the target percentage parameter is shown in Fig.~\ref{fig:kneedle}, where the algorithm identifies a critical threshold of 25\%. Once a threshold is identified, its significance is quantified using the proposed knee significance score (KSS), introduced in the following section.

\subsubsection{The Knee Significance Score (KSS): A Metric to Quantify Detection Limits}

\begin{figure}[b!]
    \centering
    \begin{subfigure}{\linewidth}
        \includegraphics[width=\linewidth]{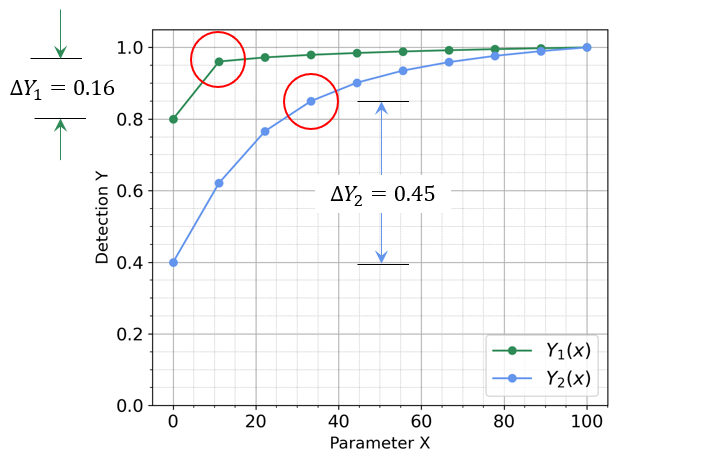}
        \caption{Raw performance drop}	
    \end{subfigure}
    
    \vspace{1em}

    \begin{subfigure}{0.745\linewidth}
        \includegraphics[width=\linewidth]{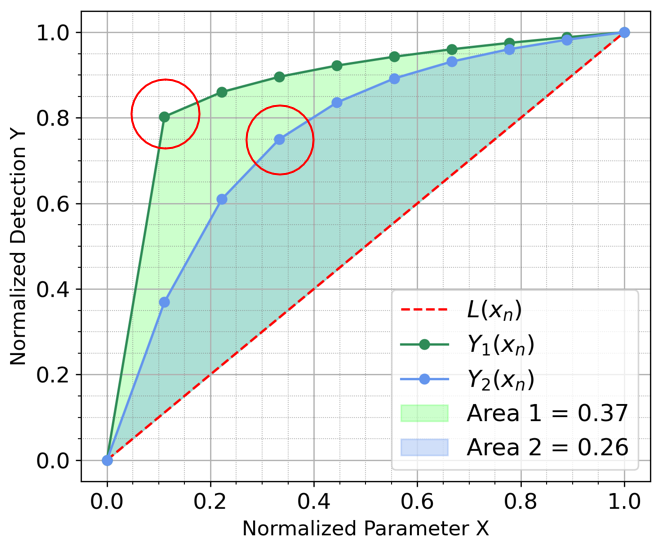}
        \caption{Normalized curvature}	
    \end{subfigure}
    
    \vspace{0.5em}
    
    \caption{Illustration of the components contributing to the knee significance score (KSS) for two detection curves: (a) raw performance drop and (b) normalized curvature. The green curve exhibits a smaller performance drop (detection impact), but larger normalized curvature (knee strength).}
    \label{fig:kss_illustration}
\end{figure}

Quantifying threshold significance is essential to distinguish meaningful transitions in detection performance from minor fluctuations. To this end, we propose a novel \textit{knee significance score} (KSS) that evaluates the relative importance of each identified knee, based on the \textit{raw performance drop} and \textit{normalized curvature} of the univariate detection curve:

{\small
\begin{align}
	\mathrm{KSS} =& \left[Y_{raw}(x_{\text{knee}})-\min (Y_{raw}(x))\right] \cdot \nonumber \\
    &\qquad \left( \frac{1}{A_{max}}\cdot \int_{0}^{1} |Y_{norm}(x)-L(x)|~dx \right)
\end{align}
}

where $Y_{raw}(x)$ is the curve in raw space, and $Y_{raw}(x_{\text{knee}})$ is the value of the identified knee point. The second term integrates the difference between the normalized curve $Y_{norm}(x)$ and the straight line connecting the endpoints, $L(x)$, normalized by the theoretical maximum area $A_{\text{max}}=1/2$.

The first term represents the raw performance drop, which represents the magnitude of performance decline at the threshold. The second term measures the normalized curvature, which quantifies how much the curve deviates from linearity (the straight line connecting endpoints). This term reflects the sharpness or abruptness of the transition, independent of the curve's overall scale. Together, these terms form a composite metric that balances the \textit{impact} on detection (raw performance drop) with the \textit{strength} of the knee (normalized curvature), providing a quantitative basis for characterizing critical transitions in detection performance.

Fig.~\ref{fig:kss_illustration} illustrates these components that contribute to the knee significance score (KSS) for two example detection curves. The knee points, highlighted with red circles, were identified by the Kneedle algorithm. The raw performance drop of $\Delta Y_1=0.16$ (green curve) and $\Delta Y_2=0.45$ (blue curve) denotes how much the wAUC detection values degrade after the knee. The normalized curvature of $\text{Area}~1=0.37$ (green curve) and $\text{Area}~2=0.26$ (blue curve) quantifies the sharpness of each curve at the threshold.

While the green curve exhibits slightly greater curvature ($\Delta~\text{Area}=0.11$), it shows a significantly smaller raw drop in detection performance ($\Delta Y=0.29$). As a result, the raw performance drop contributes more heavily to the overall KSS. Consequently, the green curve yields a lower score ($\mathrm{KSS}(Y_1)=0.12$), while the blue curve yields a higher score ($\mathrm{KSS}(Y_2)=0.23$). This comparison demonstrates how the KSS balances both the detection impact and knee strength when evaluating relative importance of parametric thresholds.

\section{Case Study to Investigate End-to-End Limitations}
\label{sec:case_study}

To demonstrate the end-to-end framework for quantifying system-level limitations, an airborne case study is conducted using seven system parameters evaluated across four canonical scenes: forest, urban, rural, and desert. These parameters reflect three core components of the remote sensing system: the scene, atmosphere, and sensor. The selected scenes span a broad range of operational scenarios to support generalizable inferences on detection limitations. The analysis is constrained to the visible and near-infrared spectral range, consistent with the characteristics of the modeled silicon-based sensor.

Baseline scenes were established following the methodology in Sec.~\ref{ssec:qualitative_scenes}, using background materials representative of typical compositions and relative abundances, as summarized in Table~\ref{tab:all_scene_fractions}. Seven manmade targets (Table~\ref{tab:mahalanobis_four_scenes}) were evaluated across the four scenes to ensure diverse target–scene combinations. FASSP input statistics were derived from atmospherically compensated reflectance data, primarily acquired during a hyperspectral mission on September 9, 2022 \cite{canas_design_2024}, and supplemented with data from an internal spectral library.

\begin{table}[h!]
    \centering
    \caption{Representative background materials and fractional abundances ($w$) used to construct baseline canonical scenes.}
    \begin{tabular}{@{}lllc@{}}
        \toprule
        \textbf{Scene} & \textbf{Class}        & \textbf{$w$ (\%)} & \textbf{Features} \\ 
        \midrule
        \multirow{4}{*}{Forest} 
        & Trees                 & 70 & Dense canopy \\ 
        & Grass                 & 15 & Ground vegetation \\ 
        & Soil                  & 10 & Forest floor \\ 
        & Stone                 & 5  & Rocks or boulders \\ 
        \midrule
        \multirow{6}{*}{Urban}  
        & Concrete              & 40 & Roads, sidewalks, parking \\ 
        & Roofing               & 20 & Building shingles \\ 
        & Trees                 & 15 & Sparse canopy \\ 
        & Grass                 & 10 & Lawns or park areas \\ 
        & Gravel                & 10 & Paths or lots \\ 
        & Soil                  & 5  & Bare earth \\ 
        \midrule
        \multirow{3}{*}{Rural}  
        & Grass                 & 60 & Open fields or roadside cover \\ 
        & Soil                  & 30 & Bare earth in paths or tilled land \\ 
        & Gravel                & 10 & Trails or unpaved roads \\ 
        \midrule
        \multirow{4}{*}{Desert} 
        & Light Soil            & 35 & Sandy or silty areas \\ 
        & Gray Soil             & 30 & Dusty, flat regions \\ 
        & Dark Gray Soil        & 25 & Rocky or moist areas \\ 
        & Vegetation            & 10 & Sparse shrubs or cacti \\ 
        \bottomrule
    \end{tabular}
    \label{tab:all_scene_fractions}
\end{table}

\subsection{System Parameters Under Investigation}

To evaluate end-to-end system behavior, this airborne case study includes three scene parameters, two atmospheric parameters, and two sensor parameters. Values were selected to represent practical ranges typically encountered in real-world scenarios. The sensitivity and limitations analyses each comprise approximately 7,000 unique parameter combinations per canonical scene (forest, urban, rural, desert), yielding roughly 28,000 scenarios. The airborne sensor was modeled at an altitude of 10 km to reflect an operational condition of an airborne sensor. All remaining system parameters in FASSP and MODTRAN were held constant and defined to represent a typical remote sensing configuration. 


\subsubsection{Scene}

Three scene parameters were selected for the case study: Mahalanobis distance (MD), target percentage (tarPerc), and t-degree of freedom (t-DOF). The Mahalanobis distance quantifies the \textit{spectral contrast} between the target reference vector and the average background distribution. The target percentage, or fill fraction, represents the proportion of a pixel occupied by the subpixel target relative to the surrounding background. Finally, the t-DOF parameter characterizes the shape of the matched filter background distribution, enabling control of scene variability, as outlined in Sec.~\ref{ssec:qualitative_scenes}.

The Mahalanobis distance between a target spectrum $\bm{x_T}$ and a multivariate background distribution $Q$ is

{\small
\begin{equation}
	\bm{d}_M(\bm{x_T},Q) = \sqrt{(\bm{x}_T-\bm{\mu})^T~\bm{\Sigma}_{ave}^{-1}~(\bm{x}_T-\bm{\mu})}
	\label{eq:mahalanobis_distance}
\end{equation}
}

where $\bm{\mu}$ is the average background vector of $Q$, and $\bm{\Sigma}_{ave}$ is the average background covariance of $Q$. In this case study, the MD serves as a parameter to characterize the spectral contrast between unique combinations of the seven manmade targets and four canonical scenes. The resulting MD values, summarized in Table~\ref{tab:mahalanobis_four_scenes}, are non-uniformly sampled due to the distinct spectral properties of each target-scene combination. Nevertheless, the selected combinations span a broad range of MD values, enabling a comprehensive analysis across varying levels of target separability.

\begin{table}[h!]
	\centering
	\caption{Mahalanobis distances between the four canonical scenes and seven manmade target materials.}
	\begin{tabular}{@{}lcccc@{}}
		\toprule
		\textbf{Target} & \textbf{Forest} & \textbf{Urban} & \textbf{Rural} & \textbf{Desert} \\
		\midrule
		Gray Felt   & 24.9  & 26.9  & 46.4  & 62.8  \\
		Gray Wood   & 25.8  & 30.3  & 46.8  & 68.0  \\
		Green Felt  & 41.8  & 36.3  & 62.7  & 41.4  \\
		Green Wood  & 61.2  & 57.3  & 87.6  & 159.1 \\
		Blue Felt   & 124.4 & 120.0 & 170.7 & 321.0 \\
		Red Wood    & 205.1 & 179.8 & 251.3 & 166.4 \\
		Red Felt    & 270.8 & 230.0 & 323.2 & 258.6 \\
		\bottomrule
	\end{tabular}
	\label{tab:mahalanobis_four_scenes}
\end{table}

\subsubsection{Atmosphere}
Two atmospheric parameters were selected for the case study: aerosol visibility (atmVis) and solar zenith angle (solZen). These parameters, modeled in MODTRAN, support a generalized representation of atmosphere conditions. Aerosol visibility (also referred to as meteorological range) describes the distance over which objects remain visible under varying levels of haze or aerosol concentration. To represent average global conditions, a mid-latitude summer model was applied in MODTRAN across all scenarios. Additionally, the relative azimuth angle between the solar and viewing planes was fixed to 0 degrees.

\subsubsection{Sensor}

Two sensor parameters were selected for the case study: viewing zenith angle (ViewZen) and the relative signal error (RSE). The RSE is an inherent parameter in FASSP that accounts for uncertainties in radiometric calibration, atmospheric compensation, and spectral mismatch. The RSE imposes a maximum signal-to-noise ratio (SNR) across all spectral bands based on the percentage relative to the maximum signal level. Given that SNR is a more interpretable metric, spectral mean SNR values were computed across all scenarios as a byproduct of the corresponding RSE values.

A sensor model was developed for the Headwall Nano-Hyperspec, with specifications representative of a typical silicon-based detector responsive to the visible and near-infrared spectral range (0.4--1 \si{\micro\meter}). Specifications were based on data sheets from the manufacturer \cite{headwall_nano-hyperspec_2016} and prior laboratory measurements \cite{conran_personal_2022}.

\subsection{Quantitative Results of Systems Case Study}

The quantitative results of this airborne case study are presented in two parts: a sensitivity analysis and a limitations analysis, following the methodologies outlined in Sec.~\ref{sec:analytical_framework}. The results demonstrate the proposed end-to-end framework and provide quantitative insights on system-level limitations in hyperspectral target detection.

\subsubsection{Sensitivity Analysis}

An overview of the seven system parameters and their corresponding values is provided in Table~\ref{tab:parameters}. These values were coarsely sampled to support generalizable insights into system-level influences on detection performance.

\begin{table}[h!]
    \centering
    \caption{Overview of seven system parameters and their corresponding values to assess sensitivities with detection.}
    \begin{tabular}{@{}llc@{}}
        \toprule
        \textbf{Stage} & \textbf{Parameter} & \textbf{Values} \\ \midrule
        \multirow{3}{*}{\centering Scene} 
        & Mahalanobis Distance &  See Table~\ref{tab:mahalanobis_four_scenes} \\ 
        & Target Percentage [\%] & [10, 20, 40, 80]  \\ 
        & t-Degree of Freedom & [2, 5, 10]        \\ \midrule
        \multirow{2}{*}{\centering Atmosphere} 
        & Aerosol Visibility [km] & [2, 10, 50]    \\ 
        & Solar Zenith [°] & [60, 40, 20]    \\ \midrule
        \multirow{2}{*}{\centering Sensor} 
        & Viewing Zenith [°] & [40, 20, 0]      \\ 
        & Relative Error [\%] & [10, 3, 0]       \\ \bottomrule
    \end{tabular}
    \label{tab:parameters}
\end{table}

\begin{figure}[t]
    \centering
    \includegraphics[width=0.85\linewidth]{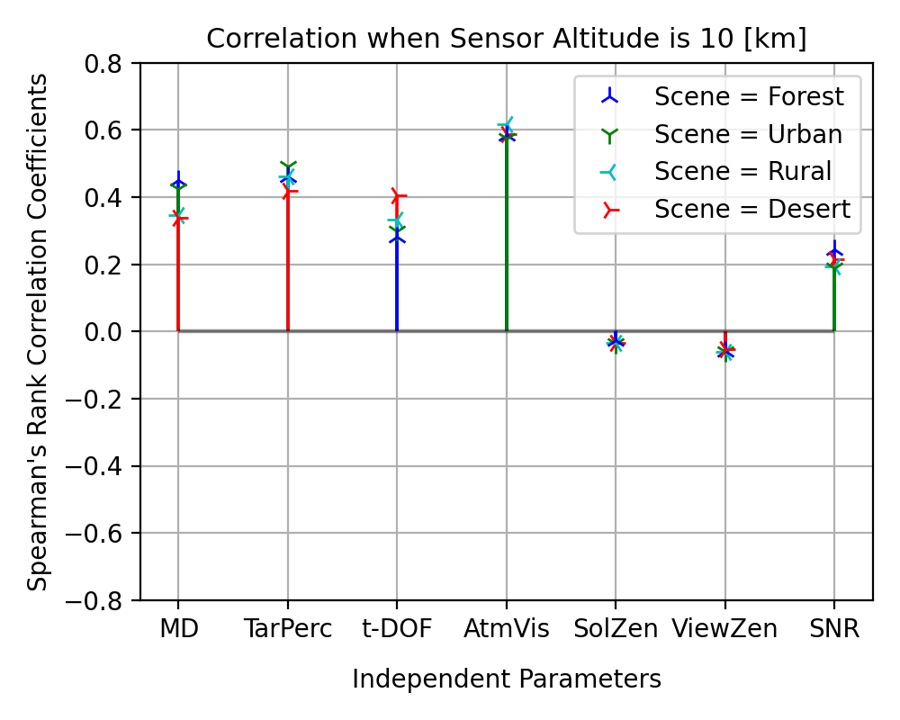}
    \caption{Correlation coefficients for an airborne altitude across four canonical scenes.}
    \label{fig:corr_coeff}
\end{figure}

To quantify system sensitivities, correlation coefficients were computed between detection performance (wAUC) and each system parameter across 6,804 unique scenarios per canonical scene (forest, urban, rural, desert). The resulting coefficients, shown in the stem plots of Fig.~\ref{fig:corr_coeff}, indicate the relative influence of each parameter on detection performance. All coefficients were statistically significant, with p-values below 0.0004, suggesting that the observed trends are unlikely to have occurred by chance.

The small variation in correlation coefficients across scenes indicates that sensitivity trends are largely scene-independent. Accordingly, dominant detection parameters are identified by averaging results across scenes. The top four system parameters are the aerosol visibility, target percentage, Mahalanobis distance, and t-degree of freedom. In contrast, the solar and viewing zenith angles rank lowest, exhibiting near-negligible influence likely due to the atmospheric compensation applied within FASSP.

\subsubsection{Limitations Analysis}

\begin{table}[h!]
	\centering
	\caption{Overview of four system parameters and their corresponding values to assess limitations of detection.}
	\begin{tabular}{@{}lc@{}}
		\toprule
		\textbf{Parameter} & \textbf{Values} \\ \midrule
		Mahalanobis Distance & See Table~\ref{tab:mahalanobis_four_scenes} \\ 
		Target Percentage [\%] & [5, 15, 25, 35, 45, 55, 65, 75, 85, 95]  \\ 
		t-Degree of Freedom & [2, 3, 4, 5, 6, 7, 8, 9, 10, 11]        \\ 
		Aerosol Visibility [km] & [1, 2, 4, 6, 8, 10, 12, 14, 16, 18]    \\ 
		\bottomrule
	\end{tabular}
	\label{tab:parameters_limits}
\end{table}

A focused limitations analysis was conducted on the top-ranked parameters identified in the sensitivity study. Each parameter was finely sampled within a narrow range (Table~\ref{tab:parameters_limits}) to precisely identify thresholds where expected detection performance sharply degrades. All lower-ranked parameters were held constant, representing a typical remote sensing configuration: solar zenith angle = $30^\circ$, viewing zenith angle = $0^\circ$, and relative sensor error = $0$.

\begin{figure*}[t]
    \centering
    \includegraphics[width=0.98\linewidth]{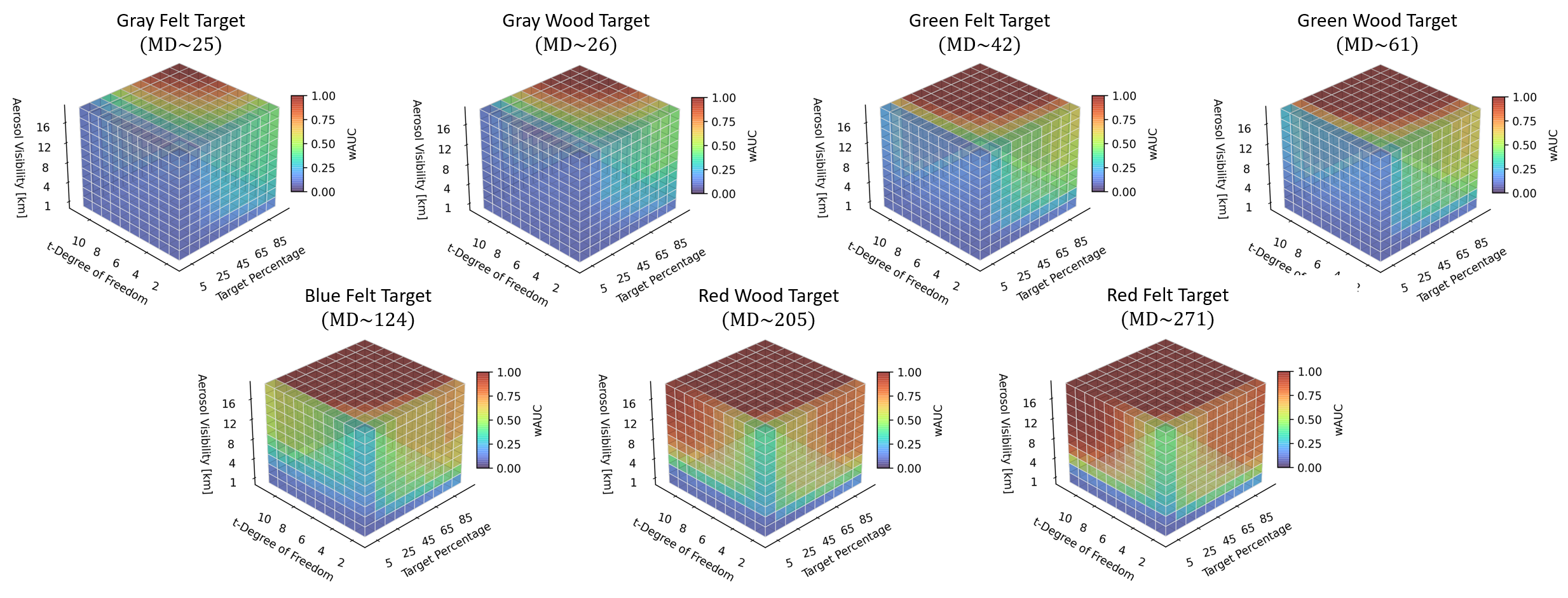}
    \caption{Detection cubes for seven manmade targets in the \textit{forest} scene, with corresponding Mahalanobis distances (MD).}
    \label{fig:detection_cubes_forest_limits}
\end{figure*}

The Mahalanobis distance is selected to function as a parameter to span a broad range of target-scene combinations. This facilitates the exploration of limitations between detection performance (wAUC) with respect to the three remaining top-ranked parameters: aerosol visibility, target percentage, and t-degree of freedom. Parametric thresholds were characterized across 7,000 unique scenarios for each canonical scene (forest, urban, rural, desert). When the four-dimensional tensor is projected onto three-dimensions, the resulting detection cubes of finely sampled parameter values can be visualized, as shown in Fig.~\ref{fig:detection_cubes_forest_limits} for the forest scene.

For each detection cube, the MD value and corresponding target provide a quantitative basis for assessing detection performance. A positive correlation between MD and wAUC values reinforces the notion that higher spectral contrast (i.e. target separability) typically improves detectability.

The non-uniform sampling of the Mahalanobis distance may bias the diversity of target-scene combinations used to infer generalizable detection limits. Specifically, the sampling is denser around low-contrast targets (i.e. gray and green) listed in Table~\ref{tab:mahalanobis_four_scenes}. To mitigate this, each four-dimensional tensor associated with a qualitative scene was linearly interpolated by resampling 10 evenly spaced MD values spanning the original range. The resulting reshaped tensors yield 10,000 unique parametric combinations, each associated with a corresponding detection performance prediction (wAUC).

\begin{figure*}[b]
    \centering
    \begin{subfigure}{0.32\linewidth}
        \includegraphics[width=\linewidth]{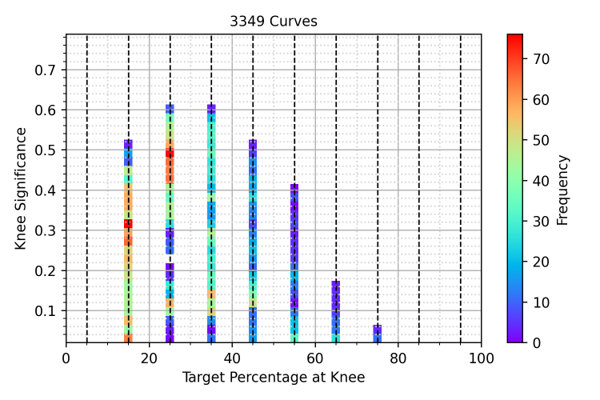}
        \caption{Target percentage}
        \label{sfig:hist_tarPerc}
    \end{subfigure}
    \begin{subfigure}{0.32\linewidth}
        \includegraphics[width=\linewidth]{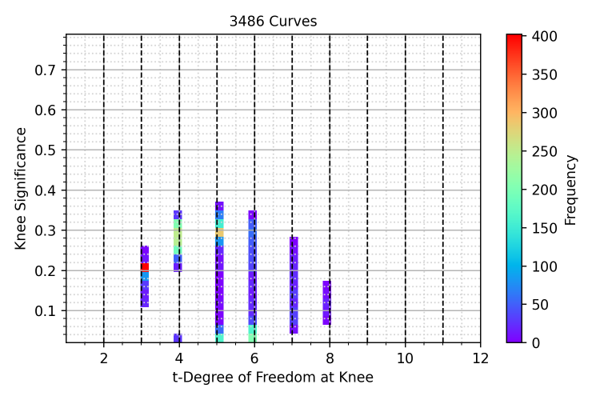}
        \caption{T-degree of freedom}
        \label{sfig:hist_tdof}
    \end{subfigure}
    \begin{subfigure}{0.32\linewidth}
        \includegraphics[width=\linewidth]{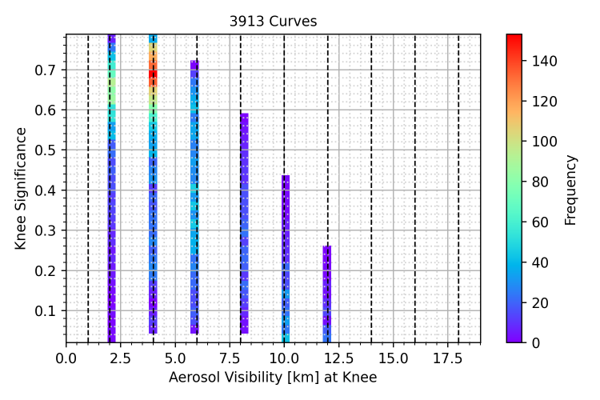}
        \caption{Aerosol visibility}
        \label{sfig:hist_atmVis}
    \end{subfigure}
    \caption{2D histograms of parametric thresholds and corresponding KSS values where detection performance is limited.}
    \label{fig:2D_hist}
\end{figure*}

Univariate detection curves were extracted from the reshaped wAUC metric tensors. To explore generalizable, scene-independent limitations, the curves were aggregated across all four canonical scenes, yielding a total of three sets of 4000 detection curves. Each set corresponds to one of the system parameters under investigation: aerosol visibility, target percentage, and t-degree of freedom.

Parametric thresholds were identified using the Kneedle algorithm across these three sets of detection curves, with knee significance scores used to quantify the importance of each detection limit. A filtering condition of KSS $<0.02$ excluded knees that were deemed insignificant. The resulting knee locations and their corresponding KSS values are visualized using 2D histograms, as shown in Fig.~\ref{fig:2D_hist}. The color intensity of each bin indicates the frequency or number of occurrences, and dashed vertical lines denote the sampled parameter values.

Detection limitations for the target percentage (Fig.~\ref{sfig:hist_tarPerc}) appear between 15--75\%, with the most frequent and significant knees (i.e. thresholds with high KSS) concentrated between 15--35\%. No knees are found at 85\%, suggesting that large subpixel target percentages do not often limit detection for cases studied. For the t-degree of freedom (Fig.~\ref{sfig:hist_tdof}), limitations arise between 3--8 degrees, with the highest concentration of significant knees around 3--5. The absences of knees at 9--11 aligns with the t-distribution's convergence toward normality at higher degrees, which reduces background clutter. Aerosol visibility (Fig.~\ref{sfig:hist_atmVis}) exhibits detection limitations primarily between 2--12 km, with the most significant knees occurring around 2--6 km and a notable concentration at 4 km. No knees are observed beyond 14 km, indicating that higher visibility does not constrain detection performance.

\section{Discussion -- Operational Implications of Systems Case Study}
\label{sec:discussion}


Comparison of the 2D histograms from the airborne case study (Fig.~\ref{fig:2D_hist}) reveals a clear hierarchy among the dominant system-level performance bottlenecks. Aerosol visibility imposes the strongest constraint, with peak knee‑significance scores (KSS~$\approx$~0.7) concentrated at visibility thresholds of 2--6 km, indicating that haze, smoke, and dust are the first‐order factors that degrade detection by reducing the target-reflected radiance reaching the sensor. Target‑percentage represents the next most significant bottleneck with peak scores (KSS~$\approx$~0.55) at 15--35\%, implying that once visibility exceeds $\approx$~6 km, the abundance of subpixel targets becomes the primary limiter on detection performance. The t‑degree‑of‑freedom exhibits the weakest influence, with modest peaks (KSS~$\approx$~0.3) at 3--5, suggesting that under clear atmospheres ($>$10 km visibility) and sufficiently large target fractions ($>$50\%), the remaining performance bottleneck is governed primarily by background clutter statistics. This hierarchy mirrors the initial sensitivity ranking obtained from the correlation‑coefficient analysis (Fig.~\ref{fig:corr_coeff}), providing independent support that the KSS‑based diagnostic is an effective tool for identifying operational detection limits. Although the analysis spans a broad range of parametric configurations, findings are bound to a VNIR sensor at an airborne altitude with specific environmental conditions established in MODTRAN. Therefore, reported thresholds and ranges should be interpreted as bottlenecks within the evaluated model and parameter space, which may not be universally applicable across all hyperspectral sensing systems and environments.

\section{Conclusion}
\label{sec:conclusion}


This research introduced an end-to-end modeling and analytical framework for quantifying system-level performance bottlenecks in optical hyperspectral target detection, providing new insight into this challenging problem. The framework enabled system trade studies across ensembles of unique scenarios using a statistical model to forecast and store traceable performance metrics within a high-dimensional tensor. Complex scenes were modeled using a parametric representation with tunable background clutter. Detection performance was evaluated using the log-weighted area under the ROC curve (wAUC), a metric well-suited for rare target detection. System-level limitations were characterized by identifying dominant parameters and associated thresholds at which detection performance degrades sharply, using a two-stage 
analysis framework. First, a sensitivity analysis based on correlation coefficients identified the most influential system parameters. Second, a limitation analysis focused on these parameters to extract critical thresholds using the Kneedle algorithm. A knee significance score (KSS) was further introduced to quantify the relative importance of thresholds across diverse scenarios.

An airborne case study was conducted using four canonical scenes (forest, urban, rural, and desert) and seven system parameters. Of these, three were scene-related, two were atmospheric, and two were sensor-specific. Sensitivity analysis of 27,216 unique scenarios identified aerosol visibility and three scene-related parameters (target percentage, Mahalanobis distance, and t-degree of freedom) as the most influential factors. A subsequent limitations analysis of 28,000 unique scenarios from the VNIR airborne sensor experiment revealed aerosol visibility (2--6 km) as the dominant performance bottleneck, followed by subpixel target percentage (15--35\%), and background clutter modeled using a t-distribution (3--5 degrees of freedom).

The proposed framework provides a multiscale approach for quantifying performance bottlenecks across diverse scene conditions and end-to-end system configurations. By identifying thresholds where detection performance sharply degrades and quantifying their relative importance, the framework supports system design, mission planning, and foundational research. Future work is recommended to explore alternative metrics for detecting performance breakpoints, extend the framework to additional spectral regimes and sensor models, and investigate higher dimensional parameter spaces. Sensor optimization studies could leverage this framework to identify channels and specifications that most strongly \textit{enable} detection, rather than merely avoiding those that \textit{limit} it. Efforts are ongoing to make this framework accessible to the broader community.

\appendices



\section*{Acknowledgment}

The authors acknowledge Philip Salvaggio and Scott D. Brown for their support in developing the C++ version of FASSP. The views presented are those of the authors and do not necessarily represent the views of DoD or its components.

\ifCLASSOPTIONcaptionsoff
  \newpage
\fi



\bibliographystyle{IEEEtran}
\bibliography{references}


%




%


\begin{IEEEbiography}[{\includegraphics[width=1in,height=1.25in,clip,keepaspectratio]{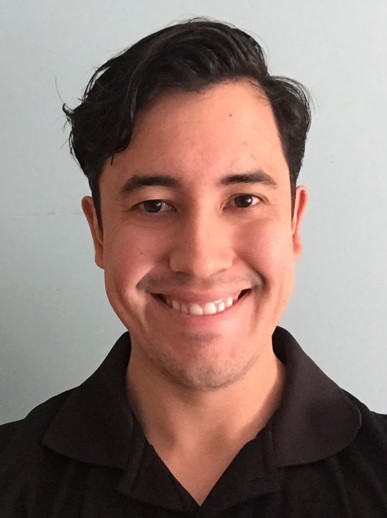}}]{Chase Ca\~{n}as} received the B.S. degree in physics from University of California, Santa Barbara, the M.S. degree in physics from California State University, Long Beach, and the Ph.D. degree in imaging science from Rochester Institute of Technology, Rochester, NY, USA.

He was with the NASA Jet Propulsion Laboratory, Pasadena, CA, USA, as an engineering graduate student from June 2019 to August 2020, where he assisted with radiometric calibration and system modeling of atmospheric sounders. From 2020 to 2025, he was a Ph.D. student with the Chester F. Carlson Center for Imaging Science, Rochester Institute of Technology, where he investigated fundamental limitations in hyperspectral target detection. His research interests include remote sensing, system modeling, sensor calibration, and image processing.
\end{IEEEbiography}


\begin{IEEEbiography}[{\includegraphics[width=1in,height=1.25in,clip,keepaspectratio]{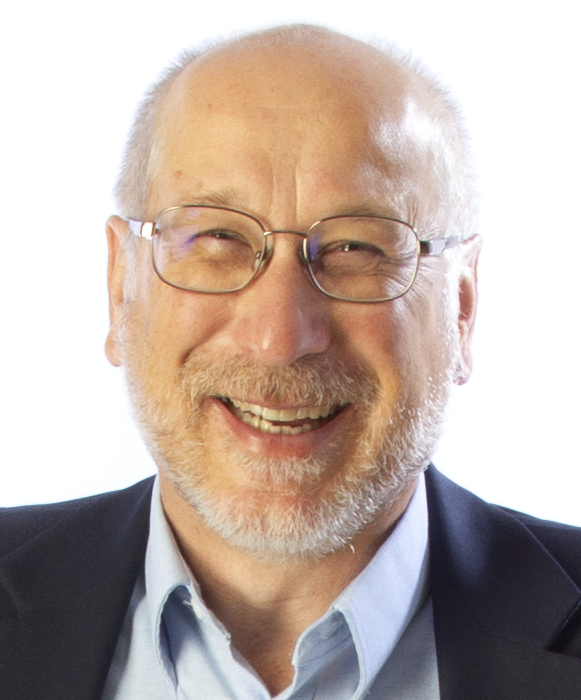}}]{John P. Kerekes} received the B.S., M.S., and Ph.D. degrees in electrical engineering from Purdue University, West Lafayette, IN, USA in 1983, 1986, and 1989.

From 1983 to 1984, he was a Member of the Technical Staff with the Space and Communications Group, Hughes Aircraft Co., El Segundo, CA, where he performed circuit design for communications satellites. From 1986 to 1989, he was a Graduate Research Assistant, working with both the School of Electrical Engineering and the Laboratory for Applications of Remote Sensing at Purdue University. From 1989 to 2004, he was a Technical Staff Member with the Lincoln Laboratory, Massachusetts Institute of Technology, Lexington, MA. Since 2004 he has been with the Chester F. Carlson Center for Imaging Science, Rochester Institute of Technology, Rochester, NY, where he is currently Research Professor.  His research interests include the modeling and analysis of remote sensing system performance in pattern recognition and geophysical parameter retrieval applications.

Dr. Kerekes is a member of Tau Beta Pi, Eta Kappa Nu, and the American Geophysical Union. He is a Senior Member of SPIE. From 1995 to 2004, he served as the founding Chair of the Boston Section Chapter of the IEEE Geoscience and Remote Sensing Society (GRSS), and from 2007 to 2010 he served as the founding Chair of the Western New York Chapter of GRSS. He was a Co-General Chair of IGARSS 2008 held in Boston, MA. From 2010 to 2015 he served on the GRSS Administrative Committee (AdCom) and from 2017 to 2024 he served as the GRSS Chief Financial Officer. In 2017 he received the GRSS Outstanding Service Award.

\end{IEEEbiography}
\vfill







\end{document}